\documentclass[12pt]{article}

\usepackage{fullpage}
\usepackage{setspace}
\usepackage{graphicx}

\usepackage[comma,numbers,sort&compress]{natbib}

\makeatletter
\renewcommand{\@biblabel}[1]{#1.}
\makeatother

\makeatletter
\def\@eqnnum{{\bf [\theequation]}}
\makeatother

\begin{document}


\begin{titlepage}
\centering

\textsf{\large
  Universal properties of quasi-one-dimensional excitons in
  semiconducting single-walled carbon nanotubes and $\pi$-conjugated polymers
}\\[15mm]

\normalsize

H. Zhao$^1$,  S. Mazumdar$^1$,  C.-X. Sheng$^2$,  and Z. V. Vardeny$^2$ \\[5mm]
$^1$Department of Physics, University of Arizona, Tucson,      AZ 85721 \\[2mm]
$^2$Department of Physics, University of Utah, Salt Lake City, UT 84112\\[15mm]

Corresponding author:\quad
Sumit Mazumdar \\[2mm]
Department of Physics, University of Arizona, Tucson, AZ 85721         \\[2mm]
Telephone: (520) 621-6803, Fax: (520) 621-4721, 
Email: sumit@physics.arizona.edu

\end{titlepage}

\onehalfspacing


\centerline{\bf Abstract}
\vskip 0.5 truein

The nature of the primary photoexcitations in semiconducting single-walled 
carbon nanotubes (S-SWCNTs) is of strong current interest. 
We have studied the emission spectra of S-SWCNTs and two
different $\pi$-conjugated polymers in solutions and films, and have
also performed ultrafast pump-probe spectroscopy on these systems.
The emission spectra relative to the absorption bands are very similar
in S-SWCNTs and polymers, with redshifted photoluminescence in
films showing exciton migration.  The transient photoinduced
absorptions (PAs) in SWCNTs and $\pi$-conjugated polymers are also
remarkably similar, with a low energy PA$_1$ and a higher energy
PA$_2$ in all cases. Theoretical calculations of excited state
absorptions within a correlated $\pi$-electron Hamiltonian find the
same excitonic energy spectrum for S-SWCNTs and $\pi$-conjugated
polymers, illustrating the universal features of quasi-one-dimensional
excitons in carbon-based $\pi$-conjugated systems.  In both cases
PA$_1$ is an excited state absorption from the optically allowed
exciton to a two-photon exciton that occurs below the continuum band
threshold. PA$_1$ therefore gives the lower limit of the binding
energy of the lowest optical exciton.  The binding energy of lowest
exciton belonging to the widest S-SWCNTs with diameters $\geq$ 1 nm in
films is 0.3--0.4 eV, as determined by both experimental and
theoretical methods.

\break

\section{Introduction}

Single-walled carbon nanotubes (SWCNTs) are of considerable current
interest because of their unique mechanical \cite{Dalton03},
electrical \cite{Collins97} and optoelectronic
\cite{Misewich03,Hertel00} properties.  Metallic versus semiconducting
character of SWCNTs are determined by their chiralities, but in all
cases extended dimensions along the tube axes and nanometer scale
diameters render these systems quasi-one-dimensional (quasi-1D).
Spatial separation of SWCNTs has recently become possible
\cite{OConnell02}, and this has led to intensive studies of the
photophysics of the semiconducting SWCNTs (S-SWCNTs). Different
experiments have begun to indicate that the primary photoexcitations
in these systems are excitons, and not the free electrons and holes
that are expected within one-electron theory \cite{SaitoBook}. Exciton
formation in S-SWCNTs is a direct consequence of the combined effects
of Coulomb electron-electron (e-e) interaction and the confinement
that occurs in 1D. It is, for example, well known that confinement
effects in 1D lead to unconditional exciton formation upon
electron-hole (e-h) excitation, and excitons in nanowires of
conventional semiconductors have been described within the 1D
hydrogenic model, with deep excitons states and discrete energy
spectrum below a Rydberg continuum \cite{Ogawa91}.
Recent theories of linear optical absorptions in S-SWCNTs have
emphasized the strong e-h interactions and the consequent exciton
formation in these systems
\cite{Ando97,Lin00,Kane03,Spataru04,Chang04,Zhao04,Perebeinos04}.

In spite of the above theoretical and experimental investigations, the
knowledge base about the physical nature of the optical excitons or
the complete excitonic energy spectrum in S-SWCNTs remains incomplete.
Whether or not the 1D hydrogenic model can be applied without
modifications to S-SWCNTs is unclear, and there is also no consensus
as yet on important materials parameters such as the exciton binding
energy. One important reason for this is that the standard technique
of comparing the thresholds of linear absorption and photoconductivity
for the determination of the exciton binding energy fails in
noncrystalline organic materials because of the existence of disorder
and inhomogeneity in these systems. More recent experimental probes of
S-SWCNTs have therefore focused on {\it nonlinear absorption}, which
can give information on the nature of excited states occurring above
the optical exciton. Several research groups have performed transient
photomodulation (PM) experiments
\cite{Lauret03,Korovyanko04,Ma04,Sheng05,Manzoni05}, which have
provided valuable information on the excitation dependence of
photoluminescence (PL) and radiative versus nonradiative relaxation
channels. A two-photon fluorescence measurement has given the first
information on the lowest two-photon state that occurs above the
lowest optical exciton in S-SWCNTs with diameters within a certain
range \cite{Wang05}. Experimental results here were interpreted within
the context of the 1D hydrogenic model. Joint experiment-theory work
on electroabsorption in S-SWCNTs has recently been performed
\cite{Kennedy05}.

It is relevant in this context to recall that there exists already a
vast literature on excitons in the $\pi$-conjugated polymers, the {\it
  other} class of carbon-based quasi-1D systems.  In order to seek
guidance from this knowledge base, we have performed precisely the
same transient PM experiments with S-SWCNTs that were previously
performed with the polymers \cite{Frolov00,Frolov01}.  We have also
performed theoretical calculations of the excited state electronic
structures and excited state absorptions for ten different S-SWCNTs
with a wide range in diameters, within the same correlated electron
model that has been widely applied to $\pi$-conjugated polymers
\cite{DGuo91,DGuo93,McWilliams91,Abe92,Beljonne97,Chandross97,Ramasesha00,Race01}.
We find a one-to-one correspondence in the excitonic energy spectra of
S-SWCNTs and $\pi$-conjugated polymers, in spite of the fact that the
carbon nanotubes are derived from 2D graphitic layers and the
coordination number per carbon atom is 3 instead of 2.

In the following sections we present the results of our experimental
and theoretical investigations.  We begin by pointing out the
similarities in the linear optical absorption and fluorescence in
S-SWCNTs and two different $\pi$-conjugated polymers.  Following this,
we present the results of our ultrafast pump-probe measurements on the
same systems. Remarkable similarities between the S-SWCNTs and
$\pi$-conjugated polymers are found.  Finally, we present our
theoretical work within the context of a correlated $\pi$-electron
Hamiltonian. We briefly review the known results for the excitonic
energy spectrum of $\pi$-conjugated polymers within the model
Hamiltonian, and then proceed to discuss our new results obtained for
the S-SWCNTs. Our main message is that there exists a {\it
  universality} in the photophysics of these two classes of systems
with quasi-1D excitons.  Our results and conclusions also suggest
possible directions for future optical investigations of S-SWCNTs.

\section{Experimental setup}

\subsection{Sample preparation}

In order to measure the ultrafast dynamic response of unbundled SWCNTs
in the mid-IR spectral range (0.1 to 1.0 eV), it was necessary to
fabricate a transparent solid sample in the broadest spectrum possible
that contains isolated nanotubes. To achieve this, 0.005\%
HiPCO-produced SWCNTs were mixed with 0.610\% SDS surfactant and
0.865\% poly-vinyl alcohol (PVA) in de-ionized water.  Sonication for
an hour before sample preparation resulted in relatively
well-separated nanotubes as evident by the sharp features in the
absorption spectrum (Fig.~1b). We then deposited a film of the
solution onto CaF$_2$ by drop-casting at 80$^\circ$C.  The film
consisted of well-separated SWCNTs embedded in an insulating matrix of
PVA having an optical density of $\sim$ 1 in the visible/near-IR
spectral range (Fig.~1).  Neither PVA nor SDS has absorption bands in
the spectral range over which we measured the absorption and transient
PM spectra. Resonant Raman scattering of the radial breathing mode was
used to determine that the nanotubes in our sample have a diameter
distribution around a mean diameter of $\sim$ 0.8 nm, and contain
about 1/3 metallic and 2/3 semiconducting SWCNTs.
\begin{figure}[htbp]
  \centering
  \includegraphics[clip]{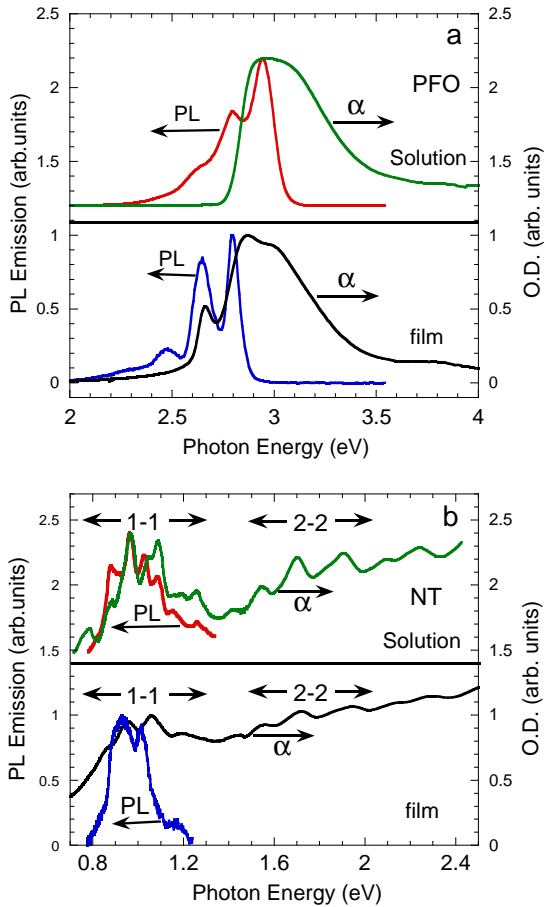}
  \caption{Photoluminescence (PL) emission and absorption
    ($\alpha(\omega)$) spectra of (a) PFO solution and film, and (b)
    isolated SWCNTs in D$_2$O solution and PVA matrix film. The
    optical transitions 1-1 and 2-2 for SWCNT are assigned.}
  \label{fig:1}
\end{figure}

Dispersions of predominantly isolated nanotubes in D$_2$O were
prepared using a procedure based on the method developed by O'Connell
{\it et al.} \cite{OConnell02}.  The sonicated samples were first
centrifuged for 10 min at 700 {\it g}.  The upper 75\% of the
supernatant was recovered using a small-bore pipette, avoiding
sediment at the bottom, and transferred to a Beckman centrifuge tube
for further centrifugation.  Samples were then centrifuged for 2 hr
at 4$^\circ$C.  The upper 50\% of the supernatant was then recovered
using a small-bore pipette, avoiding sediment at the bottom, and
transferred to a clean tube.

The semiconducting polymers used in our studies were a poly
phenylene-vinylene (PPV) derivative, viz., dioctyloxy-PPV (DOO-PPV)
that was synthesized in our laboratory using a published procedure
\cite{Hamaguchi94}, and a blue poly(9,9-dioctylfluorene) (PFO)
derivative that was purchased from American Dye Corp.~(Canada) and
used as is.  Solutions were obtained by dissolving the polymer powder
in toluene with concentration of $\sim$ 1 mg/ml; films were obtained
from the solution drop-casting onto CaF$_2$ substrates.

\subsection{The optical setup}

For our transient PM measurements we used the fs two-color pump-probe
correlation technique with linearly polarized light beams from two
different experimental setups based on Ti:sapphire lasers,
\textit{with a broad spectral range from 0.1 to 2.6 eV and 100 fs time
  resolution that was not possible before}.  To achieve such a broad
spectral range we used two laser systems: a low power high repetition
rate laser with energy per pulse of about 0.1 nJ that was used for the
mid-IR spectral range; and a high power low repetition rate laser with
energy/pulse of about 10 $\mu$J that was used in the near-IR to
visible spectral range. The transient PM spectra from the two laser
systems were normalized to each other at several probe wavelengths in
the near-IR spectral range. The ultrafast laser system used for the
low power measurements was a 100 fs Ti:sapphire oscillator (Tsunami,
Spectra-Physics) operating at a repetition rate of about 80 MHz, which
pumped an optical parametric oscillator (Opal, Spectra-Physics).  The
Opal generates signal (S) and idler (I) beams that were used as probes
with photon energy $\hbar\omega_S$ and $\hbar\omega_I$ ranging between
0.55 and 1.05 eV. In addition, these two beams were also mixed in a
nonlinear crystal (AgGaS$_2$) to generate probe at
$\omega_\mathrm{probe} = \omega_S - \omega_I$ in the spectral range of
0.13 to 0.43 eV. The pump beam for SWCNT was the fundamental at 1.6
eV; whereas for the polymers we used the second harmonic of the
fundamental at 3.2 eV. The low energy/pulse produces low
photoexcitation density of the order of 10$^{16}$ cm$^{-3}$. With such
low density we avoided the problem of exciton-exciton annihilation
that could complicate the decay dynamics, or two-photon absorption
(TPA) processes that may generate photoexcitations with very large
excess energy. To increase the signal/noise ratio, an acousto-optical
modulator operating at 85 kHz was used to modulate the pump beam
intensity. For measuring the transient response at time t with $\sim$
150 fs time resolution, the probe pulses were mechanically delayed
with respect to the pump pulses using a translation stage; the time t
= 0 was obtained by a cross-correlation between the pump and probe
pulses in a nonlinear optical crystal.  The transient PM signal,
$\Delta T/T(t)$ is the fractional change in transmission, $T$, which
is negative for photoinduced absorption (PA) and positive for
photoinduced bleaching (PB) and stimulated emission (SE).  The pump
and probe beams were carefully adjusted to get complete spatial
overlap on the film, which was kept under dynamic vacuum.  In
addition, the pump/probe beam-walk with the translation stage was
carefully monitored and the transient response was adjusted by the
beam walk measured response.

For the visible and near-IR measurements we used a home-made
Ti:sapphire laser amplifier system that operates at $\sim$1 kHz. The
laser beam was split into two beams.  The main part of the laser beam
($\sim$96\%) was used as is for pumping the SWCNT samples, or
frequency doubled in a non-linear crystal to 3.2 eV for pumping the
polymer samples. The other 4\% of the amplifier output generated white
light super-continuum pulses in a glass substrate within the spectral
range from 1.2 to 2.6 eV that was used as a probe. The pump and probe
beams were carefully adjusted to get complete spatial overlap on the
sample.

For measuring the PL spectra we used a standard cw setup comprised of
a pump laser (Ar$^+$ laser for the polymers and Ti:sapphire for the
SWCNT sample), a $\frac{1}{4}$-meter monochromator and solid-state
detectors (Si diode for the polymers and Ge diode for the SWCNT). A
phase sensitive technique was used to enhance the signal/noise ratio.
The absorption spectra were measured with commercially available
spectrometers.

\section{Experimental Results}

\subsection{Linear absorption and fluorescence}

We begin our comparison of the optical properties of $\pi$-conjugated
polymers and S-SWCNTs with a discussion of the absorption
$\alpha(\omega)$ and PL spectra of PFO and SWCNTs, in solutions and
films (see Fig.~1).  In all cases the spectra are inhomogeneously
broadened. The lowest optical gap of S-SWCNTs is in the mid-IR
spectral range, whereas that of PFO is in the blue region of the
visible spectral range.  Within the one-electron tight-binding model
\cite{SaitoBook} the 1-1 optical absorption band in SWCNTs is due to
dipole allowed transitions from the highest valence subband to the
lowest conduction subband, whereas the 2-2 transitions are from the
next highest valence subband to the next lowest conduction subband,
respectively.  It is well known that the absorption bands of M-SWCNTs
appear at $\hbar\omega > 1.6$ eV, and thus the S-SWCNTs in our films
are preferentially excited by the pump pulses at $\hbar\omega = 1.6$
eV.  In contrast to the SWCNT films, $\alpha(\omega)$ of SWCNTs in
D$_2$O solution (Fig.~1b) contains a number of distinct sub-bands.
This shows that the inhomogeneity of the SWCNTs in D$_2$O solution is
smaller than that in the films.  It is likely that in the solution the
nanotubes are bunched together in specific groups with similar
diameters, and thus exhibit more structured $\alpha(\omega)$ than in
the film.

In agreement with Kasha's rule \cite{Kasha50}, which states that light
emission occurs from the lowest energy level that is dipole-coupled to
the ground state, PL emission bands in both PFO and SWCNT samples
appear close to the low end of the $\alpha(\omega)$.  In the PFO
solution spectra (Fig.~1a) the PL 1-1 band that occurs on the high
energy side of the spectrum actually overlaps with the absorption
band. The PFO film contains two phases of different polymer order
($\alpha$ and $\gamma$), where $\gamma$ shows PL at lower energies
\cite{Cadby00}.  The PL bands of both phases are redshifted relative
to their respective absorption bands.  As in the PFO solution, in the
SWCNT solution spectrum (Fig.~1b) also the PL bands overlap with their
respective 1-1 absorption bands.  In the SWCNT film, in contrast, the
PL bands are again redshifted with respect to their corresponding
absorption bands, \textit{exactly as in the PFO film}. The underlying
mechanism of the redshifted PL band in PFO films was identified as
exciton migration to the lowest energy sites \cite{Hidayat00}.  Based
on the redshifted PL emission in the SWCNT film, we therefore
speculate that here too there occur \textit{excitons} that migrate to
the lowest energy ``emission sites.''  The ease of exciton migration
in both materials shows that the excitons are robust, with binding
energies substantially larger than the binding energies of the shallow
traps in the sample.

\subsection{Ultrafast spectroscopy measurements}

In Fig.~2 we compare the transient PM spectra of films of S-SWCNTs and
two polymers at time t = 0.  The PM spectra contain both negative (PA)
and positive (PB or SE) bands. In neither sample do we obtain a PA
spectrum that increases at low energy in the shape of free carrier
absorption, viz., $\Delta\alpha \sim \omega^{-2}$. Instead, the PA
spectra are in the form of distinct photoinduced bands PA$_1$ and
PA$_2$ associated with specific optical transitions of the primary
photoexcitations in the samples, {\it which are therefore not free
  carriers}.  The sharp PA rise below 0.2 eV is likely due to
photogenerated ``free'' carrier absorption in M-SWCNTs. The free
carriers may have been separated in the M-SWCNTs following the
original exciton photogeneration in the S-SWCNTs, and their consequent
diffusion among various SWCNTs in the sample.
\begin{figure}[htb]
  \centering
  \includegraphics[clip,scale=0.7]{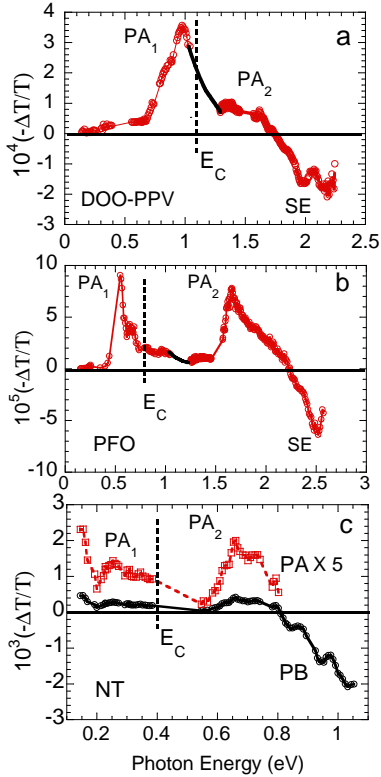}
  \caption{Transient PM spectra at t = 0 of
    films of DOO-PPV (a), PFO (b), and isolated SWCNT in PVA matrix
    (c).  Various PA, PB, and SE bands are assigned.  The vertical
    dashed lines at E$_c$ between PA$_1$ and PA$_2$ denote the
    estimated continuum band onset (see text).}
  \label{fig:2}
\end{figure}

The PA and their respective PB (or SE) bands in both polymers and
S-SWCNTs are correlated to each other (see Fig.~3). The lack of SE in
the S-SWCNT PM spectrum shows that whereas excitons in polymers are
radiative, excitons in the S-SWCNTs are not.  The dominance of
nonradiative over radiative recombination in S-SWCNT has been ascribed
to a variety of effects including, (a) trapping of the excitation at
defect sites \cite{Wang04}, (b) strong electron-phonon coupling
\cite{Htoon05}, and (c) the occurrence of optically dark excitons
below the allowed excitons \cite{Zhao04}.  Furthermore, from the
correlated dynamics of the transient PB and PA bands we have
previously concluded that PA originates from excitons in the n = 1
manifold \cite{Korovyanko04,Sheng05}.
\begin{figure}[htb]
  \centering
  \includegraphics[clip]{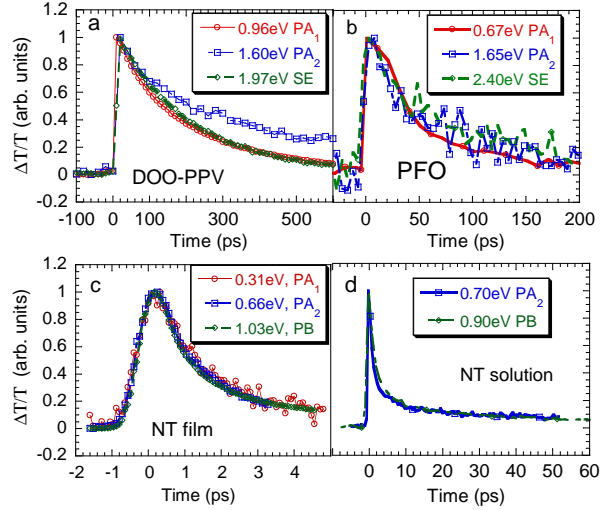}
  \caption{Transient PM dynamics at various probe energies in (a)
    DOO-PPV, (b) PFO, (c) SWCNT in PVA matrix, and (d) SWCNTs in
    D$_2$O solution.}
  \label{fig:3}
\end{figure}

Fig.~3 summarizes the decay dynamics of the various bands in the PM
spectra of these semiconductors. For each material the PA, PB or SE
bands have very similar dynamics and therefore they share a common
origin, namely, the same primary excitation, which is the
photogenerated exciton. In the case of polymer films (Fig.~3 a and b)
PA$_1$, PA$_2$ and SE decays are non-exponential, with the decay in
DOO-PPV longer than that in PFO.  It has been empirically determined
that the decay time constant $\tau$ in the $\pi$-conjugated polymers
is related to the PL quantum efficiency (QE) $\eta$ by the relation
\cite{PankoveBook}:
\begin{equation}
\eta = \tau / \tau_\mathrm{rad},
\end{equation}
where $\tau_\mathrm{rad}$ ($\approx$ 1,000 ps) is the radiative lifetime of
the 1D exciton \cite{Frolov00,Frolov01}. 
We estimate $\tau$\raisebox{-1pt}{$_\mathrm{PPV}$} = 250 ps (Fig.~3a),
            $\tau$\raisebox{-1pt}{$_\mathrm{PFO}$} = 100 ps (Fig.~3b), 
and get $\eta = 25\%$ in DOO-PPV and $\eta= 10\%$ in PFO; both
$\eta$'s are in excellent agreement with the PL QE that we measured
using an integrated sphere.

Figs.~3 c and d show the decay dynamics of the photogenerated exciton
in S-SWCNT film and solution, respectively. The decay in the S-SWCNT
film is much faster ($\sim$ 2 ps), and this is probably caused by
exciton migration in SWCNT bundles from S-SWCNTs to M-SWCNTs, the
energy relaxation rate in which is relatively fast and non-radiative.
The exciton decay in S-SWCNTs in solution (Fig.~3d) is composed of a
fast and slow component, similar to the recently measured PL(t) decay
\cite{Reich05}.  The correlation of transient PM and PL again shows
that the PA is due to excitons in this material.  We determined that
the slow component PA lasts ca.~0.5 ns.  From Eq.~1 for the QE of 1D
excitons in $\pi$-conjugated polymers and $\tau_\mathrm{rad} = 1,000$
ps we calculate $\eta \approx 50\%$ for S-SWCNTs.  The PL QE in
S-SWCNTs in solution was, however, measured to be $\eta \approx
6\times10^{-4}$, and this indicates that the radiative lifetime
$\tau_\mathrm{rad}$ for excitons in S-SWCNTs is very different than
that in polymers. Actually from the measured $\eta$ and the slow PA
component lifetime we calculate using Eq.~1 $\tau_\mathrm{rad} \approx
1$ $\mu$s, indicating that the bulk of the n = 1 excitons that
contribute to PA are nonradiative.  This seems to support the previous
suggestion that rapid decay occurs in S-SWCNTs from the optical to the
dark exciton with lower energy \cite{Zhao04}.

\section{$\pi$-electron exciton theory and photophysics}

\subsection{Theoretical model}

The striking similarities in the transient PM spectra and cw PL with
respect to $\alpha(\omega)$ spectra of the polymeric semiconductors
and the SWCNTs strongly suggest that the two families of materials
should be described within the same fundamental theory. Common to both
SWCNTs and semiconducting polymers are $\pi$-electrons, and we
anticipate that the optical behavior is determined predominantly by
these electrons.  While $\pi$-electron-only models miss the curvature
effects associated with the narrowest SWCNTs, the low peak energy of
the PA$_1$ band in Fig.~2c indicates that the photophysics of the
S-SWCNTs in our sample is dominated by the widest SWCNTs (see below).
We will thus be interested in generic consequences of e-e interactions
that are valid for the widest S-SWCNTs. We therefore focus on the
semiempirical Pariser-Parr-Pople (PPP) model Hamiltonian
\cite{PPP1,PPP3}, which has been widely applied to $\pi$-conjugated
systems in the past,
\begin{equation}
\label{ppp}
H = -\sum_{\langle ij\rangle, \sigma}t_{ij}
         (c_{i\sigma}^\dag c_{j\sigma} + c_{j\sigma}^\dag c_{i\sigma})
   + U \sum_i n_{i\uparrow}n_{i\downarrow}
   + \frac{1}{2} \sum_{i\ne j} V_{ij}(n_i-1)(n_j-1) .
\end{equation}
Here $c_{i\sigma}^\dag$ ($c_{i\sigma}$) creates (annihilates) a
$\pi$-electron on carbon atom $i$ with spin $\sigma$ ($\uparrow$,
$\downarrow$), $\langle ij\rangle$ implies nearest neighbor (n.n.)
sites $i$ and $j$, $n_{i\sigma} = c_{i\sigma}^\dag c_{i\sigma}$ is the
number of electrons with spin $\sigma$ on site $i$, and $n_i =
\sum_{\sigma}n_{i\sigma}$ is the total number of electrons on site
$i$.  The parameter $t_{ij}$ is the hopping integral between p$_z$
orbitals of n.n.~carbon atoms, $U$ is the on-site Coulomb repulsion
between two electrons occupying the same carbon atom $p_z$ orbital,
and $V_{ij}$ is the long-range intersite Coulomb interaction.  In the
case of the $\pi$-conjugated polymers $t_{ij}$ are different for
phenyl, single and double carbon bonds. For the SWCNTs, however they
are the same.  Longer range $t_{ij}$ beyond n.n.~can be included in
Eq.~2, but previous experience indicates that these terms have only
quantitative effects and do not give additional insight.

The main advantage of using the semiempirical $\pi$-electron model
over {\it ab initio} approaches is that exciton effects and excited
state absorptions can be calculated directly within the semiempirical
Hamiltonian.  The photophysics and nonlinear absorptions of
$\pi$-conjugated polymers have been widely investigated within Eq.~2
\cite{DGuo91,DGuo93,McWilliams91,Abe92,Beljonne97,Chandross97,Ramasesha00,Race01},
and very recently a theory of linear absorption in S-SWCNTs was
advanced within this model \cite{Zhao04}.  It is useful to first
briefly discuss the theory of linear and nonlinear absorptions in
$\pi$-conjugated polymers within the PPP model.

\subsection{
Excitons and excited state energy spectra of $\pi$-conjugated polymers}

Unsubstituted $\pi$-conjugated polymers usually possess inversion
symmetry and eigenstates are thus classified as even parity A$_g$ and
odd parity B$_u$.  Fig.~4a shows schematically the theoretical
excitation spectra of a light emissive $\pi$-conjugated polymer such
as PFO or PPV \cite{Chandross97}. Optical transitions corresponding to
PA$_1$ and PA$_2$ are indicated as vertical arrows in the figure.  The
spin singlet ground state is 1$^1$A$_g$. The lowest optical state
1$^1$B$_u$ is an exciton.  Although eigenstates within Eq.~2 are
correlated, the 1$^1$B$_u$ is {\it predominantly} a one electron-one
hole (1e-1h) excitation relative to the correlated ground state
\cite{Chandross99}. The lowest two-photon state, the 2$^1$A$_g$, is
highly correlated and has strong contributions from triplet-triplet
two electron-two hole (2e-2h) excitations \cite{Chandross99}. There
can occur other low energy 2e-2h triplet-triplet two-photon states
above the 2$^1$A$_g$, but all such two-photon states participate
weakly in PA or TPA, because of their weak dipole-couplings to the
1e-1h 1$^1$B$_u$.  A different higher energy two-photon state (see
Fig.~4a), referred to as the m$^1$A$_g$ (where m is an unknown quantum
number), has an unusually large dipole coupling with the 1$^1$B$_u$
and has been shown theoretically to dominate nonlinear absorption
measurements \cite{DGuo91,DGuo93,McWilliams91,Race01,Chandross99}.
The m$^1$A$_g$ is the lowest predominantly 1e-1h two-photon exciton,
and is characterized by greater e-h separation than in the 1$^1$B$_u$
\cite{Chandross99}.  The lower threshold state of the continuum band
in Fig.~4a is referred to as the n$^1$B$_u$ \cite{DGuo93,Chandross99}.
Although there exist many other excited states in the infinite
polymer, theory predicts that the 1$^1$A$_g$, 1$^1$B$_u$, m$^1$A$_g$
and n$^1$B$_u$ are the four {\it essential states}
\cite{DGuo91,DGuo93,McWilliams91,Abe92,Beljonne97,Chandross97,Ramasesha00,Race01}
that dominate the optical nonlinearity because of the very large
dipole couplings among them.
\begin{figure}[htb]
  \centering
  \includegraphics[clip]{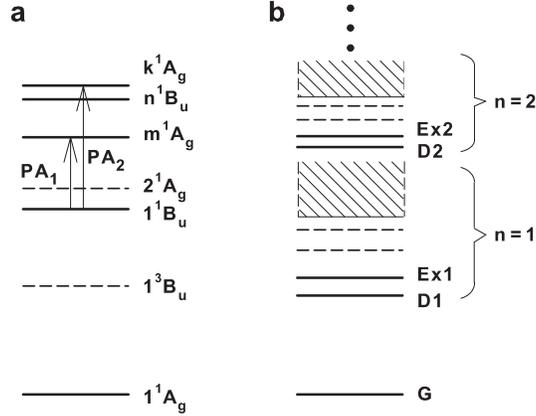}
  \caption{Schematics of the excitonic electronic structures of (a) a
    light-emissive $\pi$-conjugated polymer and (b) a S-SWCNT. In (a)
    the lowest triplet exciton 1$^3$B$_u$ occurs below the lowest
    singlet exciton 1$^1$B$_u$. The lowest two-photon state 2$^1$A$_g$
    is composed of two triplets and plays a weak role in nonlinear
    absorption.  Transient PA is from the 1$^1$B$_u$ to the m$^1$A$_g$
    two-photon exciton which occurs below the continum band threshold
    state n$^1$B$_u$, and to a high energy k$^1$A$_g$ state that
    occurs deep inside the continuum band. In (b), Ex$n$ and D$n$ are
    dipole-allowed and forbidden excitons, respectively.}
  \label{fig:4}
\end{figure}

Based on direct computations \cite{Chandross97}, PA$_1$ in PPV and PFO
corresponds to excited state absorption from the 1$^1$B$_u$ optical
exciton to the m$^1$A$_g$ two-photon exciton within the correlated
electron picture.  The m$^1$A$_g$ exciton has been experimentally
observed in PA \cite{Frolov00,Frolov01}, TPA \cite{Lawrence94},
electro-absorption \cite{DGuo93,Liess97}.  Notice that Fig.~4a
indicates that PA$_1$ gives the lower limit of the exciton binding
energy.  Given the widths of the experimental PA$_1$ bands in Fig.~2 a
and b, and the theoretical uncertainty in the energy difference
between the continuum band threshold and the m$^1$A$_g$, we believe
that the exciton binding energy is 0.8 $\pm$ 0.2 eV in PPV-DOO and
0.6 $\pm$ 0.2 eV in PFO.

PA$_2$ in $\pi$-conjugated polymers is to a distinct k$^1$A$_g$ state
that occurs deep inside the continuum band. Theoretical description of
this state has been given by Shukla {\it et al.} \cite{Shukla03}.
Specifically, in polymers with multiple bands within one-electron
theory, there occur multiple classes of 2e-2h excited configurations,
involving different bands. The exact m$^1$A$_g$ and k$^1$A$_g$ are
both superpositions of 1e-1h and 2e-2h excitations, but the natures of
the 2e-2h excitations that contribute to these states are different.

The results shown schematically in Fig.~4a have been obtained by
solving the Hamiltonian of Eq.~2 using a variety of sophisticated
many-body techniques.  Based on comparisons of exact and approximate
finite chain wavefunctions, we have shown that eigenstates that are
predominantly 1e-1h relative to the correlated ground state can be
described semiquantitatively within the single
configuration-interaction (SCI) approximation
\cite{Abe92,Chandross97}, which retains only the configuration mixing
between 1e-1h excitations from the Hartree-Fock (H-F) ground state.
We have used the SCI approximation to understand the essential states
in S-SWCNTs, which, as mentioned above, are 1e-1h.

\subsection{Linear absorption and one-photon excitons in S-SWCNTs}

We have recently calculated the linear absorptions of ten different
S-SWCNTs with diameters ($d$) ranging from 0.55 to 1.35 nm within the
PPP model of Eq.~2, using the SCI approximation \cite{Zhao04}.  The
list includes seven zigzag semiconductors between the (7,0) and the
(17,0) SWCNTS (inclusive), and the (6,2), (6,4) and (7,6) chiral
SWCNTs.  Our calculations were for the standard value of $t_{ij} = t$
= 2.4 eV. Our parametrization of the $V_{ij}$ was similar to the
standard Ohno parametrization \cite{Ohno}
\begin{equation}
\label{paramters}
V_{ij}=\frac{U}{\kappa\sqrt{1+0.6117 R_{ij}^2}} ,
\end{equation}
where R$_{ij}$ is the distance between sites $i$ and $j$ in \AA, and
$\kappa$ is a measure of the dielectric screening due to the medium
\cite{Chandross97}.  Based on our fitting of linear, nonlinear and
triplet absorptions in PPV \cite{Chandross97}, we chose $U$ = 8.0 eV
and $\kappa$ = 2 in our S-SWCNT calculations \cite{Zhao04}.

In Fig.~4b we show schematically a summary of our calculations.  There
occur multiple energy manifolds n = 1, 2, ... etc., within the total
energy scheme.  Within each n, there occur optically dark excitons
D$n$ a few k$_B$T below the optically allowed exciton Ex$n$.  Each
manifold n has also its own H-F band gap that corresponds to the lower
threshold of the continuum band within SCI theory.  The binding
energies of the excitons are then defined as the energy difference
between the H-F band gaps and the excitons within the same manifold.
We found that the binding energies of the n = 1 and n = 2 excitons
decrease with increasing diameters, and they are nearly equal in the
wide S-SWCNT limit.  Our calculated binding energies for S-SWCNTs were
in all cases smaller than those calculated for PPV or PFO using the
same parameters.  For the widest S-SWCNTs ($d\sim$ 1.3 nm), our
calculated exciton binding energies are close to 0.3 eV.

\subsection{Nonlinear absorption in S-SWCNTs}

In the present work we performed SCI calculations of excited state
absorptions using the parameters of Eq.~3, for all the S-SWCNTs in
reference \cite{Zhao04}.  The zigzag S-SWCNT calculations are for
18-20 unit cells, whereas the chiral S-SWCNT calculations are for 10
unit cells.  We have confirmed that convergence in energies has been
reached at these sizes.  Zigzag S-SWCNTs possess inversion symmetry,
and therefore eigenstates are once again classified as A$_g$ and
B$_u$.  Lack of inversion symmetry in chiral S-SWCNTs implies that
their eigenstates are not strictly one- or two-photon states.
Nevertheless, we have found from explicit calculations that even in
the chiral S-SWCNTs eigenstates are {\it predominantly} one-photon
(with negligible two-photon cross-section) and predominantly
two-photon (with very weak one-photon dipole coupling to the ground
state). We shall therefore refer to chiral S-SWCNT eigenstates as
``A$_g$'' and ``B$_u$'', respectively.

The nature of the ultrafast PA discussed in the previous section
demonstrates that PA is due to excited state absorption from the n = 1
exciton states.  From Fig.~4b, the excited state absorption can be
from Ex1, as well as from D1, following rapid nonradiative decay of
Ex1 to D1.  As in the case of $\pi$-conjugated polymers
\cite{Chandross99}, we have evaluated all transition dipole couplings
between the n = 1 exciton states (Ex1 and D1) and all higher energy
excitations.  The overall results for S-SWCNTs are very similar to
those in the $\pi$-conjugated polymers.

Our computational results are the same for all zigzag nanotubes. These
are modified somewhat for the chiral nanotubes (see below), but the
behavior of all chiral S-SWCNTs are again similar.  In Figs.~5 a and b
we show the representative results for the zigzag (10,0) and the
chiral (6,2) S-SWCNTs, respectively. The solid vertical lines in
Fig.~5a indicate the magnitudes of the normalized dipole couplings
between Ex1 in the (10,0) NT with all higher energy excitations $e_j$,
$\langle Ex1|\mu|e_j \rangle/\langle Ex1|\mu|G \rangle$, where $G$ is
the ground state.  The dotted vertical lines are the normalized
transition dipole moments between the dark exciton D1 and the higher
excited states, $\langle D1|\mu|e_j \rangle/\langle Ex1|\mu|G
\rangle$.  Both couplings are shown against the quantum numbers $j$ of
the final state along the lower horizontal axis, while the energies of
the states $j$ are indicated on the upper horizontal axis. The reason
why only two vertical lines appear in Fig.~5a is that all other
normalized dipole couplings are {\it invisible on the scale of the
  figure}.  A striking aspect of the results for the (10,0) zigzag
S-SWCNT are then that {\it exactly as in the $\pi$-conjugated
  polymers, the optical exciton Ex1 is strongly dipole-coupled to a
  single higher energy m$^1$A$_g$ state}.  The dark exciton D1 is
similarly strongly coupled to a single higher energy state (hereafter
the m$'^1$A$_g$). Furthermore, the dipole couplings between Ex1 and
m$^1$A$_g$ (or D1 and m$'^1$A$_g$) are stronger than those between the
ground state and the excitons, which is also true for the
$\pi$-conjugated polymers \cite{Chandross99}.
\begin{figure}[htb]
  \centering
  \includegraphics[clip]{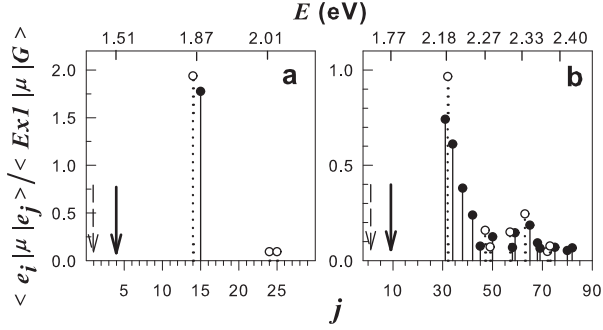}
  \caption{Normalized transition dipole moments between S-SWCNT exciton states
    Ex1 and D1 and all other excited states $e_j$, where $j$ is the
    quntum number of the state in the total space of single
    excitations from the H-F ground state. The numbers along the upper
    horizontal axes are energies in eV. Results shown are for (a) the
    (10,0), and (b) the (6,2) S-SWCNTs, respectively. Solid (dotted)
    lines correspond to $e_i$ = Ex1 (D1).  The solid and dashed arrows
    denote the quantum numbers of Ex1 and D1, respectively.}
  \label{fig:5}
\end{figure}

The situation in the chiral (6,2) S-SWCNT is slightly different, as
shown in Fig.~5b.  Both the Ex1 and D1 excitons are now strongly
dipole-coupled to several close-lying excited states, which form
narrow ``bands'' of m``$^1$A$_g$'' and m$'$``$^1$A$_g$'' states.
Similar to the case of zigzag S-SWCNTs, these bands occur above the
Ex1.

From the calculated results of Fig.~5 a simple interpretation to
PA$_1$ in Fig.~2 emerges, viz., PA$_1$ is a superposition of excited
state absorptions from Ex1 and D1.  This raises the question whether
PA$_2$ in the S-SWCNTs can be higher energy inter-subband absorptions
from the n = 1 excitons to two-photon states that lie in the n = 2 (or
even n = 3) manifolds.  We have eliminated this possibility from
explicit calculations: the transition dipole matrix elements between
one-photon states in the n = 1 manifold and two-photon states within
the higher n manifolds are zero.  As in the $\pi$-conjugated polymers,
two-photon states giving rise to PA$_2$ cannot therefore be
computationally accessed without taking into account the 2e-2h
excitations \cite{Shukla03} and are outside the scope of the present
work.

In Fig.~6 we show the energy locations of all the relevant one- and
two-photon states within the n = 1 manifold for the (10,0) and (6,2)
NTs. The figure includes the absolute energies of the Ex$n$ and D$n$
excitons, the two-photon m$^1$A$_g$ states and the corresponding H-F
bandgaps for the (10,0) and (6,2) S-SWCNTs.  We did not include in
Fig.~6 the m$'^1$A$_g$ states that are coupled to the dark excitons,
as they will be indistinguishable in their energies from the
m$^1$A$_g$ states on this scale.  For comparison, we have also
included in the figure the excitonic energy spectrum of PPV,
calculated within the PPP Hamiltonian of Eq.~2 with the same
parameters.  Within SCI theory, all states below the H-F thresholds
within the same manifold are excitons.  For the (10,0) zigzag S-SWCNT
we have also shown the higher energy n$^1$B$_u$ state in the n = 1
manifold. As with the $\pi$-conjugated polymers, this state is
identified by its large dipole couplings with the m$^1$A$_g$.
Importantly, in both the (10,0) and the (6,2) S-SWCNTs, the m$^1$A$_g$
and m``$^1$A$_g$'' states occur below the respective H-F thresholds,
indicating that exactly as in the $\pi$-conjugated polymers, the
energy locations of these excitonic two-photon states give the lower
bound to the exciton binding energy.
\begin{figure}[htb]
  \centering
  \includegraphics[clip]{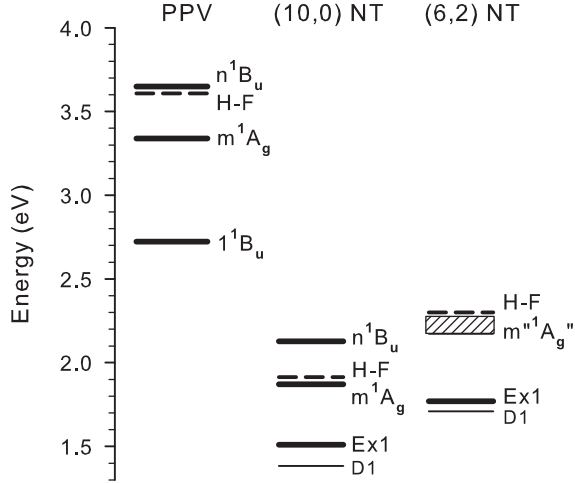}
  \caption{SCI energies of optically relevant states in (from left to right) PPV,
    (10,0) SWCNT and (6,2) SWCNT, respectively. In all cases the
    m$^1$A$_g$ (m``$^1$A$_g$'') is an exciton and PA$_1$ gives the
    lower limit of the exciton binding energy.}
  \label{fig:6}
\end{figure}

In Table 1 we have summarized the calculated results for all the
S-SWCNTs we have studied.  We give the calculated PA energies that
originate from both the Ex1 and the D1 excitons, as well as the
corresponding H-F thresholds for the n = 1 manifold. Note that for
S-SWCNTs with diameters 0.8--1.3 nm, the calculated range of PA
energies, $\sim$ 0.25--0.55 eV, matches closely with the experimental
width of the PA$_1$ band in Fig.~2.  As we have already indicated,
PA$_2$ is not due to excited state absorption from the n = 1 excitons
to states in the higher manifold, and this indicates that the most
likely origin of PA$_2$ is the same as that in PPV. In Fig.~2 we have
indicated that the continuum band threshold in PPV and PFO lies in
between PA$_1$ and PA$_2$, in view of both theoretical and
experimental work.  Based on the overall experimental similarities
that we find between the $\pi$-conjugated polymers and S-SWCNTs, and
the theoretical energy spectra, it is then natural to guess that the
location of the n = 1 continuum band threshold in S-SWCNTs also lies
between the peaks of the PA$_1$ and PA$_2$ absorption bands. This
would give lowest exciton binding energies of about 0.3 eV in the
widest S-SWCNTs, in excellent agreement with the calculations.
\begin{table}[htb]
  \centering
 \begin{tabular}{ccccc}
\hline\hline
NT & $d$ (\AA) & $E_{PA_1}^{Ex1}$ (eV) & $E_{PA_1}^{D1}$ (eV) & $E_{b1}$ (eV) \\ \hline
 (7,0) & 5.56 & 0.42             & 0.54 & 0.54 \\
 (6,2) & 5.72 & 0.41 $\sim$ 0.44 & 0.47 & 0.53 \\
 (8,0) & 6.35 & 0.37             & 0.47 & 0.53 \\
 (6,4) & 6.92 & 0.36             & 0.41 & 0.48 \\
(10,0) & 7.94 & 0.36             & 0.49 & 0.41 \\
(11,0) & 8.73 & 0.31             & 0.40 & 0.41 \\
 (7,6) & 8.95 & 0.28 $\sim$ 0.33 & 0.38 & 0.39 \\
(13,0) & 10.3 & 0.31             & 0.43 & 0.32 \\
(14,0) & 11.1 & 0.28             & 0.37 & 0.34 \\
(17,0) & 13.5 & 0.27             & 0.35 & 0.29 \\ \hline \hline
\end{tabular} 
  \caption{Summary of computed results for different SWCNTs. Here
$E^{Ex1}_{PA_1}$ and $E^{D1}_{PA_1}$ are PA$_1$ energies that
correspond to excited state absorptions from the Ex1 and D1,
respectively.  $E_{b1}$ is the binding energy of Ex1, as measured by
the energy difference between the H-F band threshold and the energy of
Ex1.}
  \label{tab}
\end{table}

\section{Discussions and conclusions}

Our principal conclusion is that the energy spectrum within the n = 1
energy manifold of S-SWCNTs is very similar to the energy spectrum of
$\pi$-conjugated polymers.  The origin of the low energy PA$_1$ in
both S-SWCNTs and $\pi$-conjugated polymers is then excited state
absorption from EX1 and D1 to higher energy two-photon excitons. The
broad nature of the PA$_1$ band in the S-SWCNTs arises from the
inhomogeneous nature of the experimental sample, with SWCNT bundles
that contain a distribution of S-SWCNTs with different diameters and
exciton binding energies. If we assume that the peak in the PA$_1$
band corresponds to those S-SWCNTs that dominate nonlinear absorption,
then the low energy of the peak in the PA$_1$ band in Fig.~2, taken
together with the data in Table 1, suggest that PA is dominated by the
widest S-SWCNTs in our sample. The common origin of PA$_1$ and PA$_2$
(see Fig.~3) indicates that PA$_2$ is also dominated by the widest
S-SWCNTs. The peak in the PA$_2$ band at $\sim$ 0.7 eV then is due to
the widest S-SWCNTs, with PA$_2$ due to narrower S-SWCNTs occurring at
even higher energies.  Hence the energy region 0.2--0.55 eV in
Fig.~2c must correspond only to PA$_1$ excitations.  Based on the
similarities in the energy spectra of the S-SWCNTs and the
$\pi$-conjugated polymers in Fig.~6, we can therefore construct the
vertical dashed line in Fig.~2c, which identifies the threshold of
the continuum band for the widest S-SWCNTs in the film. Exciton
binding energies of ca. 0.4 eV are then predicted for those S-SWCNTs
in the film that dominate the nonlinear absorption.  For S-SWCNTs with
diameters $\sim$ 0.8 nm, Wang {\it et al.} found TPA at energies very
close to the peak of our PA$_1$ band, and based on additional
estimates from the 1D hydrogenic exciton theory, determined their
exciton binding energies to be also ca.~0.4 eV \cite{Wang05}.
Remarkably, our calculated binding energies of the excitons for
S-SWCNTs with diameters in this range are very close in Table 1.  We
believe that this coincidence in the calculated and observed
experimental exciton binding energies for the S-SWCNTs is not
fortuitous.  Within the PPP Hamiltonian the exciton binding energy
depends primarily on the long range Coulomb interactions in Eq.~3.
The parametrization of Eq.~3 for PPV was arrived at following an
extensive search across fifteen sets of ($U, \kappa$) values in
reference \cite{Chandross97}, and only the parameter set $U$ = 8.0 eV,
$\kappa$ = 2.0 reproduced all four linear absorption bands (at 2.4,
3.7, 4.7 and 6.0 eV, respectively) and the experimentally observed
energies of the m$^1$A$_g$ \cite{Frolov00,Frolov01} and the 1$^3$B$_u$
\cite{Monkman01}.  The successful transferability of the parameters
from PPV to S-SWCNTs then suggests that the Coulomb interaction
parameters as well as the background dielectric constants in these two
classes of materials are close in magnitude. This in turn justifies
the use of $\pi$-electron models for the S-SWCNTs, at least for the
widest S-SWCNTs which dominate the nonlinear absorption.

The similarity in the photophysics of S-SWCNTs and $\pi$-conjugated
polymers suggests several directions for future research.
Theoretically, determination of the proper mechanism of PA$_2$ in
analogy to the existing results for $\pi$-conjugated polymers
\cite{Shukla03} is a high priority.  The energy location of the lowest
triplet exciton and triplet PA energy in S-SWCNTs are also of strong
experimental and theoretical interest.  The energy difference between
the lowest singlet and triplet excitons is a measure of the strength
of Coulomb interactions, while calculations of triplet PA provide yet
another check for the model Hamiltonian of Eq.~2. S-SWCNTs also
provide us with a unique opportunity of extending nonlinear
spectroscopy to the very interesting region of the n = 2 energy
manifolds.  Exciton states in the n = 2 manifold lie deep inside the n
= 1 continuum band and can be expected to exhibit novel optical
behavior. Whether or not the n = 2 manifold has the same structure as
the n = 1 manifold is an intriguing question.  Although neither TPA
nor PM experiments can reach the two-photon states in this region,
these should become visible in electro-absorption measurements.
Theoretical and experimental works along these directions are
currently in progress.


\section{Acknowledgments}
We thank Ray H. Baughman and Alan B. Dalton
(Nanotech Institute, UT Dallas) for supplying the SWCNT samples, and M. Tong 
for help with the ps measurements in the visible spectral range.
Work at the University of Arizona was supported by NSF-DMR-0406604. 
Work at the University of Utah    was supported by DOE FG-04-ER46109.

\newpage
\singlespacing

\end{document}